Revisiting $LaMnO_3$: A density functional theory study

Ju Hyeon Lee and Bongjae Kim*

*Department of Physics, Kyungpook National University, Daegu 41566, Republic of Korea*

*E-mail: bongjae@knu.ac.kr

Abstract

Density functional theory has been widely applied to a variety of realistic materials but often fails to accurately describe the properties of correlated systems. The DFT + *U* method, which introduces a Hubbard *U* correction to the DFT, has been instrumental in the study of systems such as transition metal oxides. However, the methodological details of DFT + *U* and its specific impact on the electronic structure and magnetic properties of correlated systems remain incompletely understood. In this study, using the prototypical transition metal oxide $LaMnO_3$ as an example, we systematically assess the performance of two distinct DFT + *U* methods: spin-polarized DFT + *U* (SDFT + *U*) and spin-unpolarized charge-only DFT + *U* (CDFT + *U*). We found that, while the Coulomb *U* acts similarly in both approaches, the Hund's $J_H$ plays a fundamentally different role, particularly in determining the correct magnetic phases. Our investigation demonstrates the active role of $J_H$ in the exchange splitting, leading to distinct magnetic ground states based on the different schemes. We further investigate the associated magnetic exchange interactions and compare our results with beyond-DFT methods.

## 1. Introduction

Density functional theory (DFT) is a well-established theoretical framework for describing the many-body electrons in solid-state systems and has been very successful in predicting the ground-state electronic structures of materials. However, in DFT, the many-body electron motions are treated within an effective single-particle limit, and there is a fundamental shortcomings [1,2]. Hence, for systems where the many-body electron–electron interactions are significant, DFT based on simple local density approximation (LDA) and generalized gradient approximation (GGA) often fail to provide a correct ground phase, and beyond-DFT methods are required [3,4]. Improved methodologies include hybrid functionals, the GW approximation, DFT + dynamical mean-field theory, and DFT + *U*. Among these, DFT + *U* has been widely adopted due to its transparent formulation, where the Hubbard *U* parameter is added to the one-body functional, and its relatively low computational cost compared to more demanding approaches. Consequently, DFT + *U* is widely used for investigating correlated materials such as transition metal compounds [1,5,6].

$LaMnO_3$ (LMO) is a representative transition metal oxide that has been extensively studied using various DFT-based theoretical methods. Upon systematic doping, the system undergoes a metal–insulator transition accompanied by distinct magnetic phases, which are intimately linked to the changes in the electronic structure [7,8]. At optimal doping level, the system exhibits colossal magnetoresistance (CMR), which has attracted considerable attention for its potential applications [9]. In its undoped form, the parent compound LMO crystallizes in an orthorhombic symmetry with a *Pnma* symmetry [10] [Fig. 1] and shows an insulating ground state with an A-type antiferromagnetic (A-AFM) order [Fig. 1(c)].

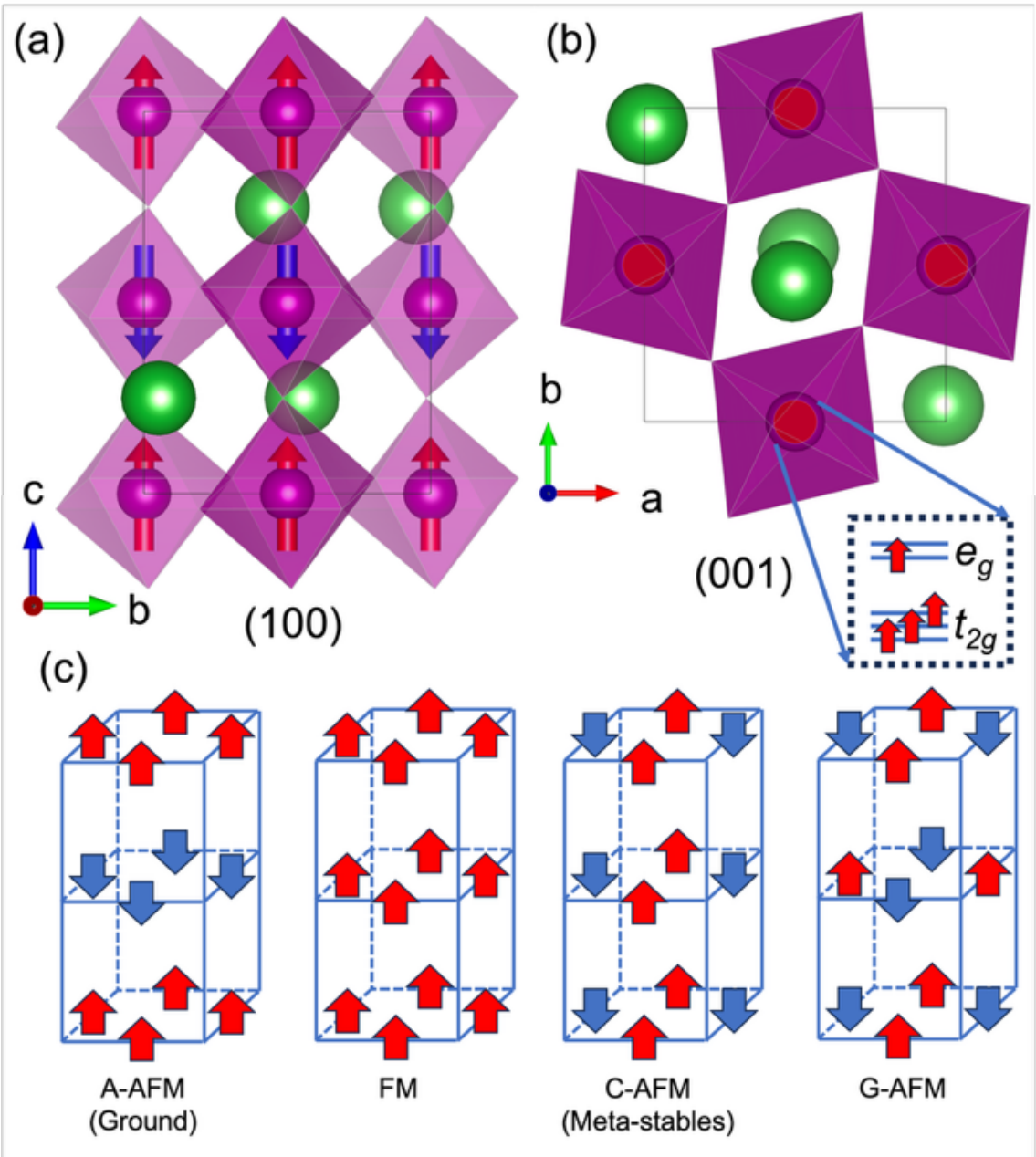

**Fig. 1.** Experimentally reported crystal structure of $LaMnO_3$ shown along the (a) (100) and (b) (001) directions. The purple octahedra and green spheres represent the $MnO_6$ octahedra and La atoms, respectively. The dashed box indicates the local electron occupation for $Mn^{3+}$ state. (c) Possible magnetic configurations of $LaMnO_3$ are presented, including the ground-state magnetic order, A-AFM (A-type antiferromagnetic), and metastable states: FM (ferromagnetic), C-AFM (C-type antiferromagnetic), and G-AFM (G-type antiferromagnetic). Red and blue arrows denote opposite directions of the local spin moments at the Mn sites.

Previous studies have demonstrated the failure of standard DFT approaches, such as LDA and GGA, in predicting the correct A-AFM ordering in LMO [11,12]. To address this issue, subsequent studies employed beyond-DFT approaches, such as hybrid functionals and DFT + $U$ schemes [13,14]. Particularly, DFT + $U$ studies have underlined the importance of different implementations of the Hubbard $U$ in DFT and discussed the explicit role of the Hund's $J_H$ [15,16].

In this study, we present a comprehensive investigation of LMO within the framework of DFT with special focus on DFT + $U$ methods. We compare two different implementations: spin-polarized DFT + $U$ + $J_H$ (SDFT + $U$) and spin-unpolarized charge-only DFT + $U$ + $J_H$ (CDFT + $U$). We analyze the electronic structures and magnetic ground states obtained from two methods and explore roles of the Coulomb $U$ and Hund's $J_H$. Our $U$–$J_H$ phase diagrams demonstrate clear differences between the two approaches. Furthermore, we map the energetics of various magnetic configurations onto the Heisenberg Hamiltonian and extract magnetic exchange interactions. We then discuss the underlying origins of the different responses of magnetic exchanges to variations in the Hubbard parameters. In addition, we examine various exchange–correlation functionals of different levels of approximation.

## 2. Methods

All calculations were performed using the Vienna *ab initio* Simulation Package (VASP) [17]. For the exchange–correlation functional, the LDA and two types of GGA functionals, Perdew–Burke–Ernzerhof (PBE) [18] and PBE revised for solids (PBEsol) [19] were employed. We also included the strongly constrained and appropriately normed (SCAN) meta-GGA functional [20] and the Heyd–Scuseria–Ernzerhof (HSE06) hybrid functional [21]. Plane-wave energy cutoff of 600 eV and a 6×6×4 Monkhorst–Pack k-point mesh were used.

We compared two different implementations of DFT + *U*: SDFT + *U* [22] and CDFT + *U* [23]. In SDFT + *U*, the exchange–correlation energy depends on the spin-resolved charge densities, whereas in CDFT + *U*, it depends only on the total charge density. Accordingly, the charge density functionals in the two schemes differ: in SDFT + *U*, the Hubbard *U* term is added on top of the spin-polarized charge functional, while in CDFT + *U*, the spin-unpolarized charge functional is used. Such differences lead to different treatment of the double-counting correction. The Hubbard *U* is applied to the correlated Mn–3*d* subspace in both schemes, with spin degrees of freedom included. The double-counting energy is expressed as follows:

$$E_{\mathrm{dc}}^{\mathrm{SDFT}} = \frac{U}{2}n(n-1) - \frac{J_H}{2}\sum_{\sigma}^{\uparrow,\downarrow} n^{\sigma}(n^{\sigma}-1) \quad (1)$$

$$E_{dc}^{CDFT} = \frac{U}{2}n(n-1) - \frac{J_H}{2}n\left(\frac{n}{2}-1\right) \quad (2)$$

here, $E_{dc}^{SDFT(CDFT)}$ denotes the double-counting energy in the spin-(un)polarized functional, $n$ is the occupation number (density matrix), and $\sigma$ is the spin channel. Differences in the double counting correction, hence different energy corrections, separates two DFT + *U* schemes. Also, the role of Hund's $J_H$ differs significantly between the two schemes as we will discuss in detail below. We further compared our results with Dudarev's implementation, which uses a rotationally invariant DFT + $U_{\mathrm{eff}}$ formalism with $U_{\mathrm{eff}} = U - J_H$, such that Hund's $J_H$ is not explicitly included [24]. We found that CDFT scheme yields a slightly smaller volume compared with SDFT scheme. However, to avoid additional complications arising from the structural degree of freedom, we adopted the experimental structures for all our calculations, as reported in ref. [10].

## 3. Results

### 3.1. Magnetic phase diagram

First, we discuss the phase diagrams obtained using two distinct DFT + $U$ schemes. In Fig. 2 presents the ground-state magnetic phase diagrams for four different exchange-correlation functionals—LDA, PBE, PBEsol, and SCAN—constructed as based on the two Hubbard interaction parameters: Coulomb $U$ and Hund's $J_H$. The phase diagram reveals the fundamental differences between the two DFT + $U$ schemes.

In the case of SDFT + $U$ scheme, the A-AFM ground state is stabilized in the small-$U$ and small-$J_H$ regime [Fig. 2(a)]. As $U$ and $J_H$ increase, the magnetic configuration evolves in distinct ways. Increasing the Coulomb $U$ drives the system toward a ferromagnetic (FM) ordering, while an increase in Hund's $J_H$ stabilizes the G-type antiferromagnetic (G-AFM) ordering. Within a simple spin model considering next-nearest-neighbor Mn sites, the A-AFM configuration is composed of four in-plane ferromagnetic and two out-of-plane antiferromagnetic exchange interactions [Fig. 1]. As $U$ increases, the out-of-plane Mn–Mn interactions favor ferromagnetism, causing a transition from A-AFM to FM. In contrast, increasing $J_H$ progressively stabilizes antiferromagnetic interactions between nearest-neighbor Mn sites, resulting in a transition from FM through A-AFM to G-AFM. Notably, we found that the magnetic phase diagrams do not vary significantly upon the different functionals. The respective roles of $U$ and $J_H$ remain consistent across the various functionals within SDFT + $U$ framework.

In contrast, the magnetic phase diagram obtained using the CDFT + $U$ scheme is markedly different. In this case, the system exhibits a G-AFM ground state for small $U$ and $J_H$ regime. Increasing either $U$ or $J_H$ promotes FM interactions between Mn sites, leading to a transition into an A-AFM phase, in which in-plane interactions become ferromagnetic. With a further increase in $J_H$, the out-of-plane exchange also becomes ferromagnetic, eventually stabilizing the FM ground state. As we discuss in detail below, the response of the electronic structure to Hund's $J_H$ differs significantly between the two schemes, which accounts for the distinct phase diagrams. Importantly, the phase space exhibiting the correct magnetic structure, A-AFM, is significantly narrower in the CDFT + $U$ case--except for SCAN functional--than in the SDFT + $U$ case, illustrating the nontrivial magnetic behavior of LMO. This highlights the challenges in the accurate description of the magnetism of LMO within DFT framework [11-14]. In general, the choice of exchange-correlation functional does not drastically modify the CDFT + $U$ phase diagrams. However, a notable exception is found in the SCAN functional, where the magnetic phase diagram exhibits an almost no dependence on $U$, a point that will be further discussed in a later section.

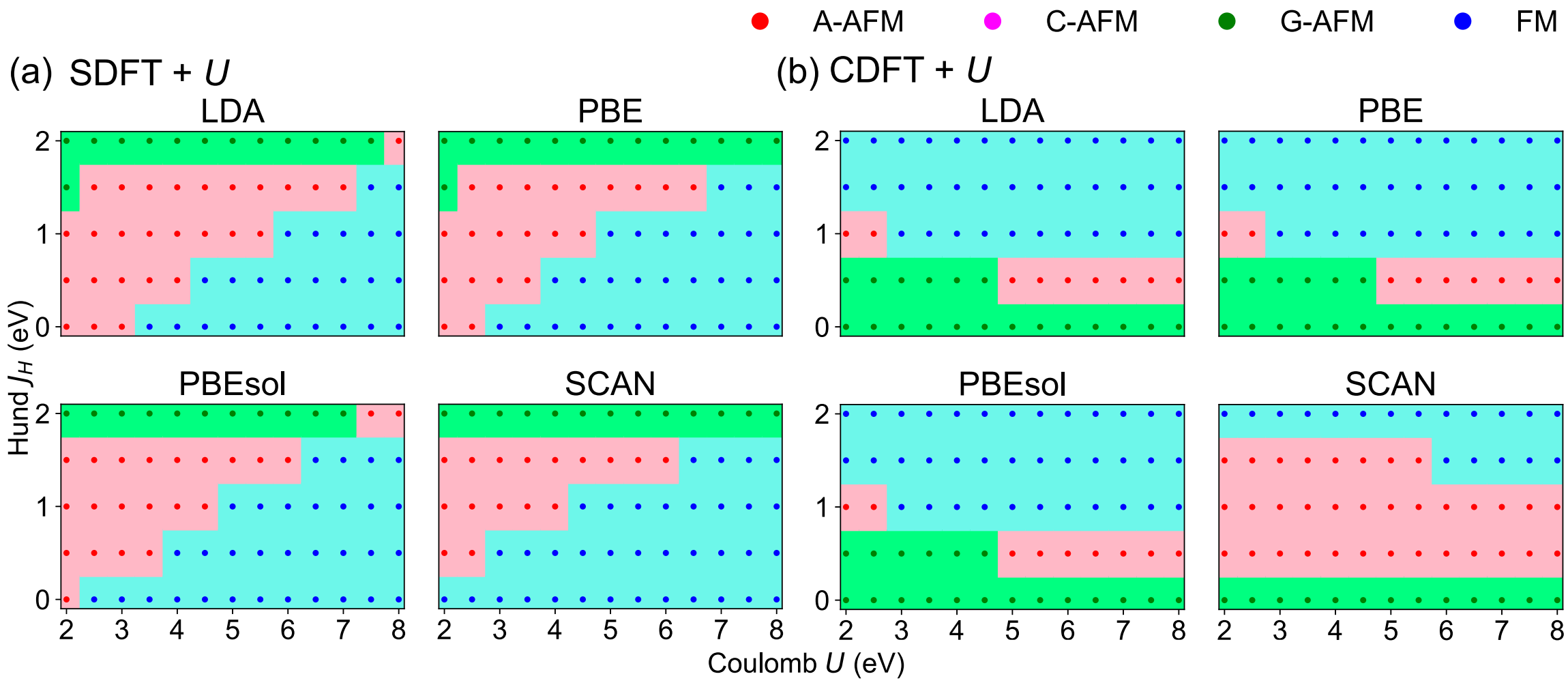

**Fig. 2.** Ground-state magnetic phase diagrams as functions of the Coulomb *U* and Hund's $J_H$ for (a) SDFT + *U* and (b) CDFT + *U* schemes. The magnetic configurations correspond to those shown in Fig. 1(c).

### 3.2. Electronic structure

Before investigating the magnetic exchange interactions, we first examine the evolution of the electronic structure, as it is closely connected to the magnetic properties. We compare how the electronic structure evolves as a function of Coulomb *U* and Hund's $J_H$ for the two DFT + *U* schemes. From a local ionic perspective, the $Mn^{3+}$ ion in $LaMnO_3$ has four *d*-electrons. Three electrons fully occupy the $t_{2g}$ states, and one electron half-fills the $e_g$ state in the majority spin channel for a high-spin configuration. A *d*-*d* gap, more specifically, $e_g$–$e_g$ gap, opens due to Mott-type correlation effects. The Jahn–Teller distortion further splits the $e_g$ states, thereby increasing the gap size.

In Fig. 3, we systematically vary *U* while fixing $J_H$ = 1 eV. As *U* increases, the unoccupied Mn-3*d* orbitals in both spin channels shift to higher energies, enlarging the Mott-type gap in both the SDFT + *U* and CDFT + *U* schemes [14]. While there are minor differences in the projected density of states (PDOS), the overall electronic structure exhibits a similar trend: the unoccupied orbitals, particularly in the minority spin channel, move to higher energy levels. The partially filled majority $e_g$ orbital progressively splits with increasing *U*. In principle, the Coulomb *U* is applied to all *d*-orbitals, but partially filled states are more sensitive and show more pronounced response. This behavior is consistent across both SDFT + *U* and CDFT + *U* methods. Notably, even though the ground-state magnetic configuration changes with *U* [Fig. 2], the PDOS does not vary significantly between different magnetic phases. All presented PDOS data correspond to the A-AFM magnetic configuration.

We next examine the effect of Hund's $J_H$ on the electronic structure. Even at the DFT level without *U*, the two DFT + *U* schemes show inherent differences: SDFT solves separate Kohn–Sham equations for each spin channel, whereas CDFT treats only the total charge density. As a result, at $J_H$ = 0 eV, the two schemes already show notable differences. When $U$ is applied on top of the spin-polarized DFT solution in the SDFT + $U$ scheme, a sizable spin-dependent gap emerges. In contrast, CDFT + $U$ scheme produces minimal spin splitting. These differences are clearly visible in the $J_H$ = 0 eV cases shown in Fig. 4, where $U$ = 6 eV is used. Here, we note that the overall physics remains consistent for Fig. 3 and Fig. 4 across the reasonable ranges of Hubbard and Hund parameters used in this study.

As Hund's $J_H$ increases, the two schemes how markedly contrasting behaviors. In the SDFT + $U$ method, the spin splitting is already pronounced, and the role of $J_H$, which is expected to increase the total spin moment, is rather limited. Moreover, increasing $J_H$ effectively reduces the strength of the Coulomb interaction through the renormalization $U_{\text{eff}} = U - J_H$. This behavior is recapped with Dudarev's simplified $U_{\text{eff}}$ implementation, in which $J_H$ is not treated explicitly [14]. As shown in Fig. 5, the PDOS from the SDFT + $U_{\text{eff}}$ scheme closely resembles that from the SDFT + $U$ scheme with $U_{\text{eff}}$, confirming that $J_H$ acts primarily to reduce the effective Coulomb repulsion without influencing the spin splitting.

In contrast, the CDFT + $U$ scheme shows an active role of Hund's $J_H$. According to Hund's first rule, $J_H$ enhances the total spin $S$ of the system. In our case, increasing $J_H$ leads to a pronounced increase in the spin splitting between majority and minority $t_{2g}$ states. Notably, the energy splitting in the $t_{2g}$ channel reaches values significantly larger than the applied atomic $J_H$ (< 1 eV) scale. Although there is some renormalization of the effective $U$ near the Fermi level, particularly within the $e_g$ states, the dominant contribution in the electronic structure change arise from Hund's $J_H$, which enhances the splitting. This behavior is markedly different to the case of SDFT + $U$.

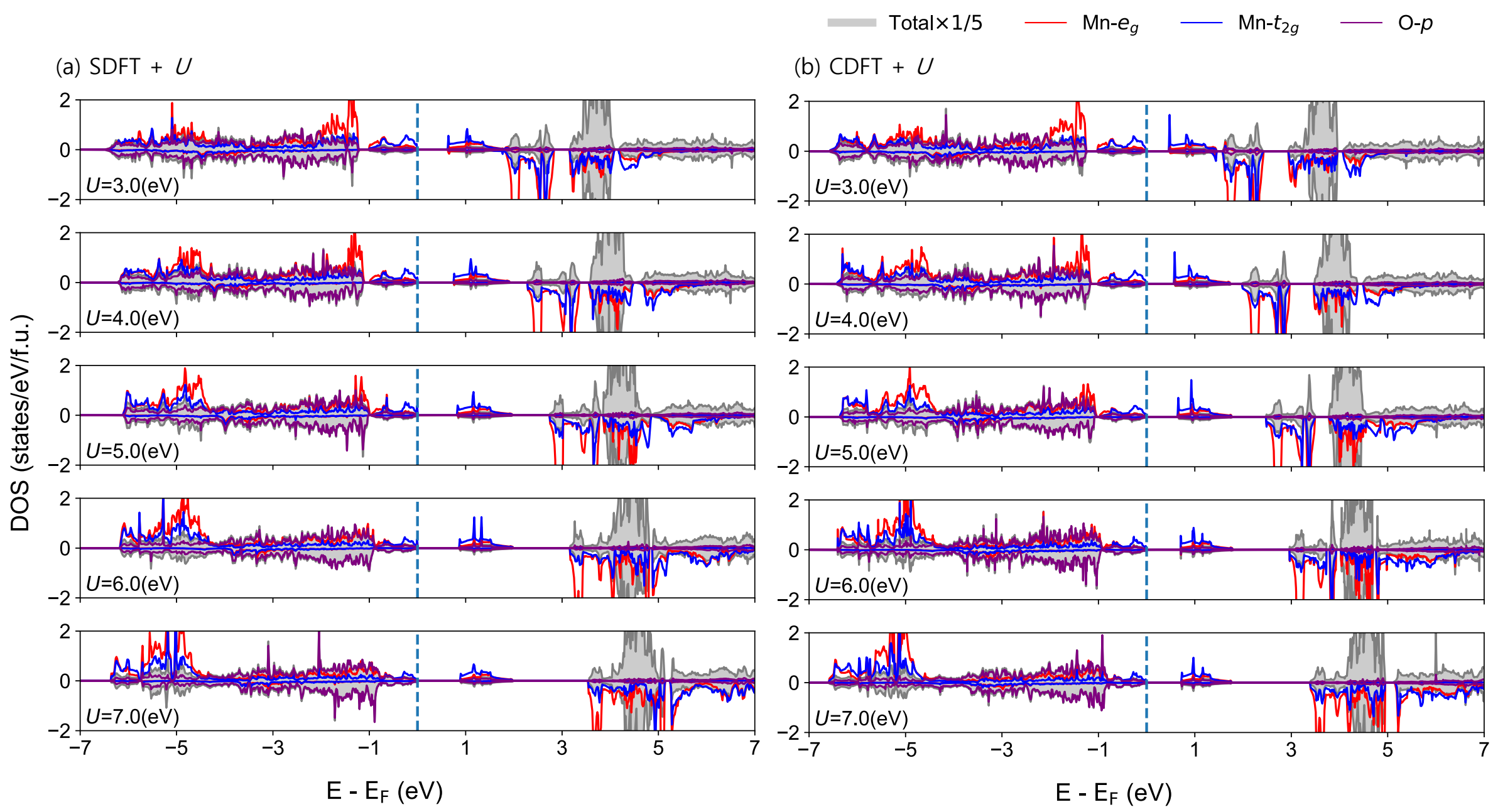

**Fig. 3.** Projected density of states calculated using the PBE functional is shown for SDFT + $U$ (a) and CDFT + $U$ (b). The Hund's $J_H$ was fixed at 1 eV. Red and blue lines represent the Mn-$e_g$ and $t_{2g}$ orbitals, respectively, while the O-$p$ orbital is shown in purple. The total density of states is indicated by gray shading and scaled by a factor of 1/5 for clarity.

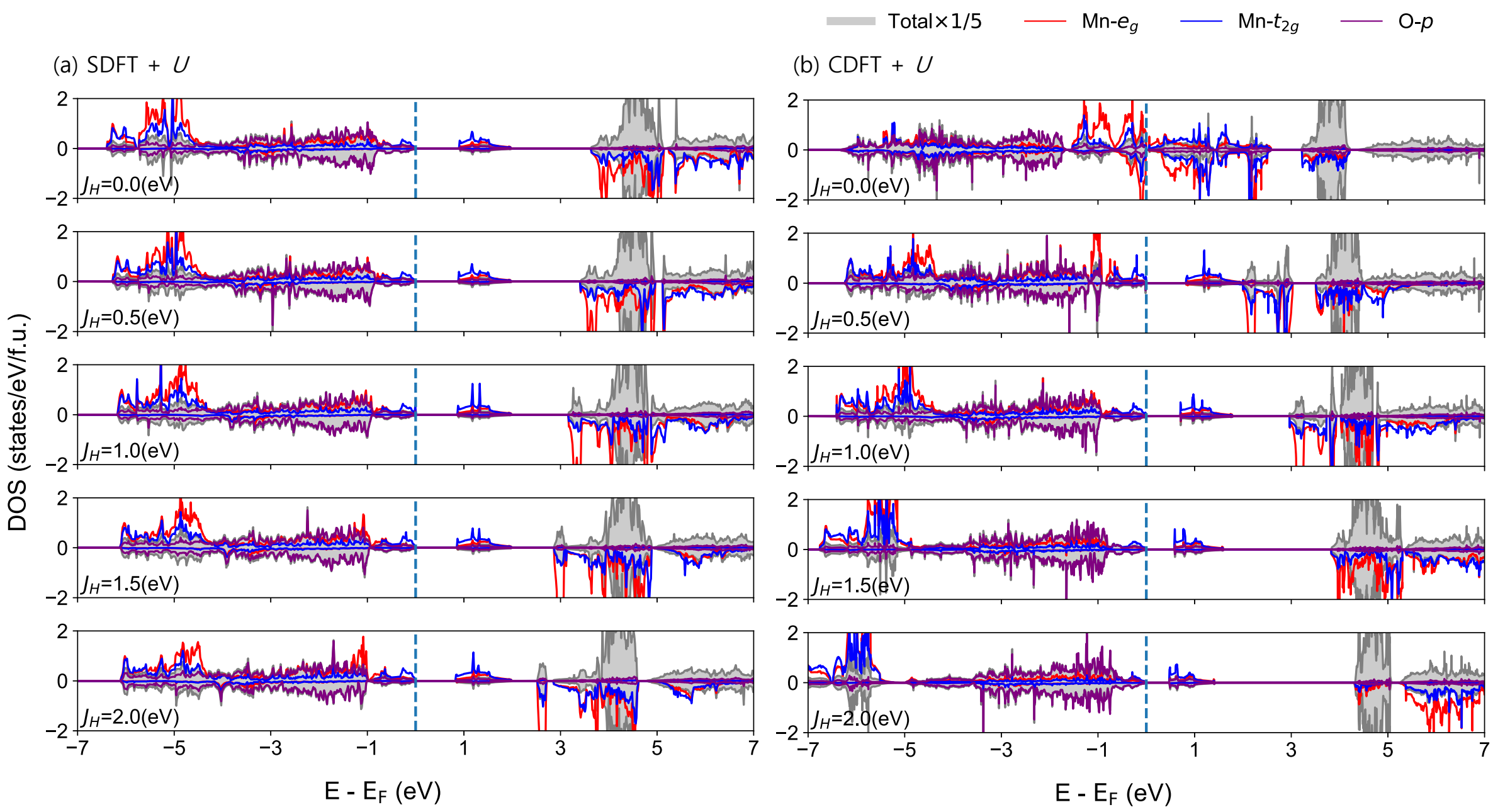

**Fig. 4.** Projected density of states calculated using the PBE functional for SDFT + $U$ (a) and CDFT + $U$ (b). The Coulomb interaction parameter $U$ was fixed at 6 eV. The color coding is the same as in Fig. 3.

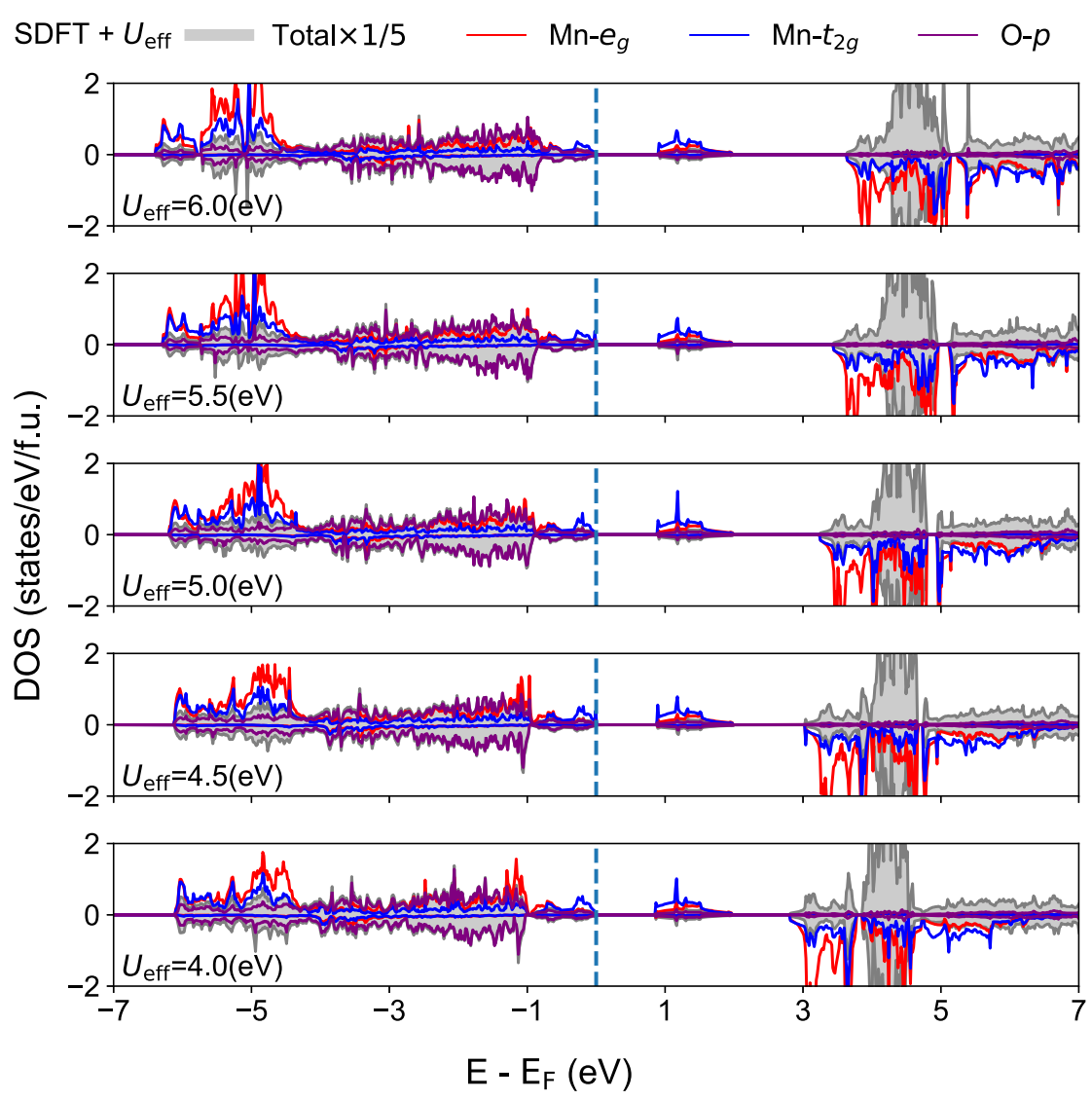

**Fig. 5.** Projected density of states calculated within the SDFT + $U_{eff}$ scheme ($U_{eff} = U - J_H$) using the PBE functional. The Coulomb interaction parameter $U$ was fixed at 6 eV. The color coding is the same as in Fig. 3. The occupied electronic structure is comparable to that obtained from the SDFT + $U$ scheme shown in Fig. 4(a).

### 3.3. Magnetic exchange couplings

To gain a microscopic understanding of the diverse magnetic phases in the $U$-$J_H$ phase diagram, we investigate the leading magnetic exchange couplings in the system. Based on the total energies of different magnetic configurations shown in Fig. 1, we map the energetics of magnetic orders onto a Heisenberg-type spin Hamiltonian, as follows:

$$H_{spin} = \frac{1}{2}\sum_{\langle ij\rangle} J_{ij} S_i S_j \tag{3}$$

where, $J_{ij}$ denotes the magnetic exchange parameter between site $i$ and $j$, and $S_i$ is the classical spin at the Mn $i$-site. As the magnetic moment originates from localized $d$-electrons, we employ a minimal model that includes only nearest-neighbor couplings. These interactions are categorized into in-plane ($J_{ab}$) and out-of-plane ($J_c$) ones (within the $ab$ plane and along the $c$ axis, respectively; see Fig. 1). In this model, the complexity of magnetic exchanges, including orbital-dependent contributions between two Mn sites, is absorbed into two effective parameters, $J_{ab}$ and $J_c$. The exchange parameters are extracted from the energies of different magnetic configurations:

$$J_{ab} = \frac{E_{C-AFM} - E_{FM}}{64}\ (J_{ab}^{C-F}) = \frac{E_{G-AFM} - E_{A-AFM}}{64}\,(J_{ab}^{G-A}) \tag{4}$$

$$\boldsymbol{J_c} = \frac{\boldsymbol{E_{A-AFM} - E_{FM}}}{\boldsymbol{32}}\,(\boldsymbol{J_c^{A-F}}) = \frac{\boldsymbol{E_{G-AFM} - E_{C-AFM}}}{\boldsymbol{32}}\ (\boldsymbol{J_c^{G-C}}) \tag{5}$$

As shown, each exchange parameter can be evaluated using two different sets of magnetic configurations. For example, $J_{ab}$ can be derived either from the energy difference between C-AFM and FM or between G-AFM and A-AFM, as shown in Eq. (4). Similarly, $J_c$ is derived from the combinations shown in Eq. (5). We attempt to understand the $U$-$J_H$ phase diagram in Fig. 2 from exchange couplings.

Fig. 6 shows the exchange couplings as functions of Coulomb $U$ for both the SDFT + $U$ and CDFT + $U$ schemes. We employ the PBE functional and note that the functional dependence is minor, except for SCAN, as can be seen in the magnetic phase diagram [Fig. 2]. For both schemes, Coulomb $U$ enhances the FM tendency for both $J_{ab}$ and $J_c$, consistent with the phase diagram in Fig. 2. As $U$ increases, regardless of $J_H$, the interaction strengths generally become more negative, indicating stronger FM interaction. The sign of the exchange parameters change at around the phase boundaries, which aligns well with the transitions observed in Fig. 2. In the SDFT + $U$ case, we observe a crossover of $J_c$ from positive to negative near $U$ = 2.5 eV for $J_H$ = 0 eV (see red curve), which corresponds exactly to the A-AFM to FM transition in Fig. 2. The critical $U$ for the $J_c$ sign change increases systematically with $J_H$, mirroring the trend seen in the SDFT + $U$ phase diagram. In contrast, for CDFT + $U$, when $J_H$ = 0 eV [Fig. 4(b)], the positive $J_{ab}$ (blue and black curves) and $J_c$ (red curve), which support the G-AFM phase, decrease but remain positive up to $U$ = 8 eV. The system remains FM across the entire range, consistent with Fig. 2. For $J_H$ = 0.5 eV, the sign of $J_{ab}$ changes to negative at around $U$ = 4.5 eV, and one obtains an A-AFM phase. $J_c$ also decreases but remains positive, so the system is A-AFM even for large $U$. For $J_H$ = 1.0 eV, $J_{ab}$ is negative throughout, and $J_c$ becomes negative at around $U$ = 2.5 eV, resulting in a transition to G-AFM. For even higher $J_H$, the system remains as FM in the CDFT + $U$ scheme.

As presented in Fig. 2, the contrasting roles of $J_H$ are further confirmed in our analysis of exchange interactions. In the SDFT + $U$, increasing $J_H$ enhances AFM tendencies, as evident from the overall positive sign of the exchanges. This is the opposite for CDFT + $U$, where increasing $J_H$ drives the exchanges toward negative values, favoring FM.

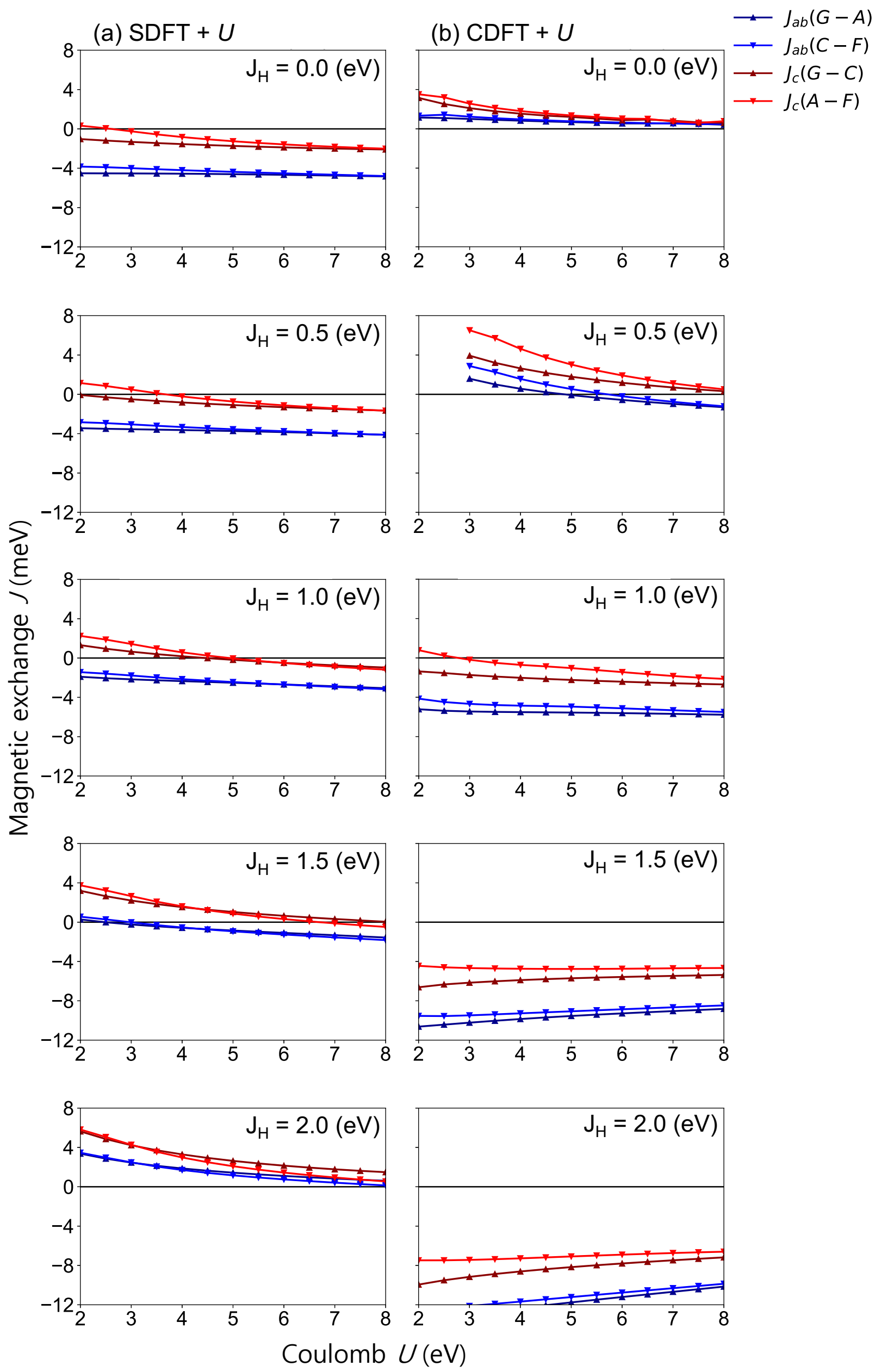

**Fig. 6.** Magnetic exchange parameters calculated from different magnetic configurations using the PBE functional for (a) SDFT + $U$ and (b) CDFT + $U$ schemes. Ferromagnetic interactions correspond

to negative exchange values ($J < 0$), while antiferromagnetic interactions correspond to positive values ($J > 0$).

In fact, the sign of the magnetic exchanges arises from the complex orbital-orbital interactions. In an octahedral crystal field, the Mn $d^4$ configuration favors FM interactions primarily through half-filled $e_g$-$e_g$ exchanges, which may overcome AFM contributions from $t_{2g}$- $t_{2g}$ channels. However, in LMO, strong Jahn-Teller distortions split the $e_g$ orbitals, intensifying competition, especially along the interlayer Mn sites [25]. In our model [Eq. (3)], all such interactions are incorporated into site-based effective exchange parameters.

Generally, increasing $U$ enhances the splitting between majority and minority spin channels, thereby weakening AFM interactions that are dominated by $t_{2g}$–$t_{2g}$ coupling. In contrast, the $e_g$-$e_g$ gap in the majority spin channel remains relatively unchanged, leaving FM interactions largely unaffected. This can explain the overall AFM-to-FM evolution trend in Fig. 6. In the SDFT + $U$ scheme, this behavior is consistent across varying $J_H$, which effectively reduces $U$ via $U_{\text{eff}} = U - J_H$. In CDFT + $U$, this FM tendency is also observed at small $J_H$; however, as $J_H$ increases, the response of the exchange coupling reverses. For instance, at $J_H$ = 2.0 eV, the exchange becomes weaker with increasing $U$, indicating a reduction in FM tendency. Furthermore, in CDFT + $U$, $J_H$ actively enhances FM behavior, which is contrasted to the AFM enhancement observed in SDFT + $U$.

This behavior is from the different roles played by $J_H$ in each scheme. Unlike SDFT + $U$, where $J_H$ effectively renormalizes $U$, CDFT + $U$ exhibits more active role for $J_H$. In CDFT + $U$, exchange splitting between majority and minority $d$-bands is small; even at $U$ = 6 eV, without $J_H$, the Mn $d$-states remain in a low-spin configuration [Fig. 4(b)]. Upon inclusion of $J_H$, we observe a clear increase in exchange splitting, highlighting the active role played by Hund's $J_H$ in CDFT + $U$ [15,16].

Noteworthy is that the exchange parameters obtained from two different magnetic configuration sets [Eqs. (4) and (5)] reveal notable differences between the SDFT + $U$ and CDFT + $U$ schemes. That is, $J_{ab}$ and $J_c$ derived from different configurations (e.g., C-AFM and FM, or G-AFM and A-AFM) yield slightly different values. This discrepancy is more pronounced in CDFT + $U$, likely due to our oversimplified nearest-neighbor Heisenberg model [Eq. (3)], which neglects the longer-range interactions [26]. This difference indirectly highlights the non-local nature of magnetic exchange in LMO.

To examine the locality from the electronic structure point of view, we show the spin-polarized partial charge density in Fig. 7. The two cases show very similar charge profiles in real space, which imply that this discrepancy stems not from the locality, but rather from complex exchange processes involving states away from the Fermi level.

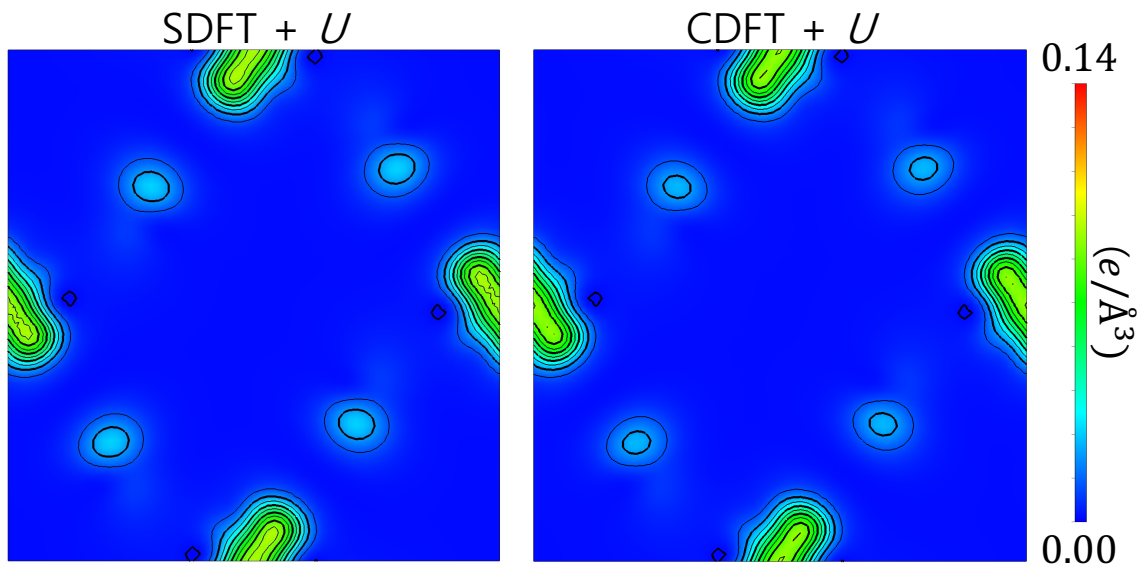

**Fig. 7.** The spin-polarized charge density is plotted over the energy range from 0 eV to −1 eV, with the Fermi energy set to 0 eV. The Coulomb interaction *U* was set to 6 eV, and Hund's $J_H$ was set to 1 eV.

### 3.4. SCAN and HSE approaches

Finally, we compare the overall electronic structure and magnetic phases using the meta-GGA SCAN functional and the hybrid functional HSE06 scheme. The effectiveness of the SCAN functional has been well demonstrated in previous studies [27,28], and Varignon *et al.* explicitly showed its improved description of the electronic, magnetic, and lattice properties in transition metal oxide perovskites without the inclusion of Hubbard parameters [29]. For LMO, however, we found that the SCAN functional still requires proper values of *U* and $J_H$ for the correct description of magnetic phases, as shown in Fig. 2. When comparing the electronic structures of different functionals without the inclusion of *U*, the SCAN functional shows enhanced characterization of the electronic structure. As shown in Fig. 8, compared to the LDA, PBE, and PBEsol functionals, SCAN better captures the spin splitting of the system, which resembles the inclusion of the Hubbard *U* correction [Fig. 3(b)]. Nevertheless, sizable Hubbard parameters are still needed for a more quantitatively accurate description.

The phase diagram for SCAN is similar to those of other functionals in the SDFT + *U* scheme, but it differs markedly for the CDFT + *U* scheme. In this case, we observe significantly extended A-AFM are in the phase diagram. When employing sizable Hund's $J_H$ from 0.5 to 1.5 eV, the correct magnetic configuration is consistently found over a much wider range of *U* values. This extended regime of the correct magnetic ground state can be interpreted as improved performance of the SCAN functional. However, there is ongoing discussion about the fundamental capability of SCAN to accurately describe magnetic properties in such systems, and further clarification is anticipated from future studies [30-32].

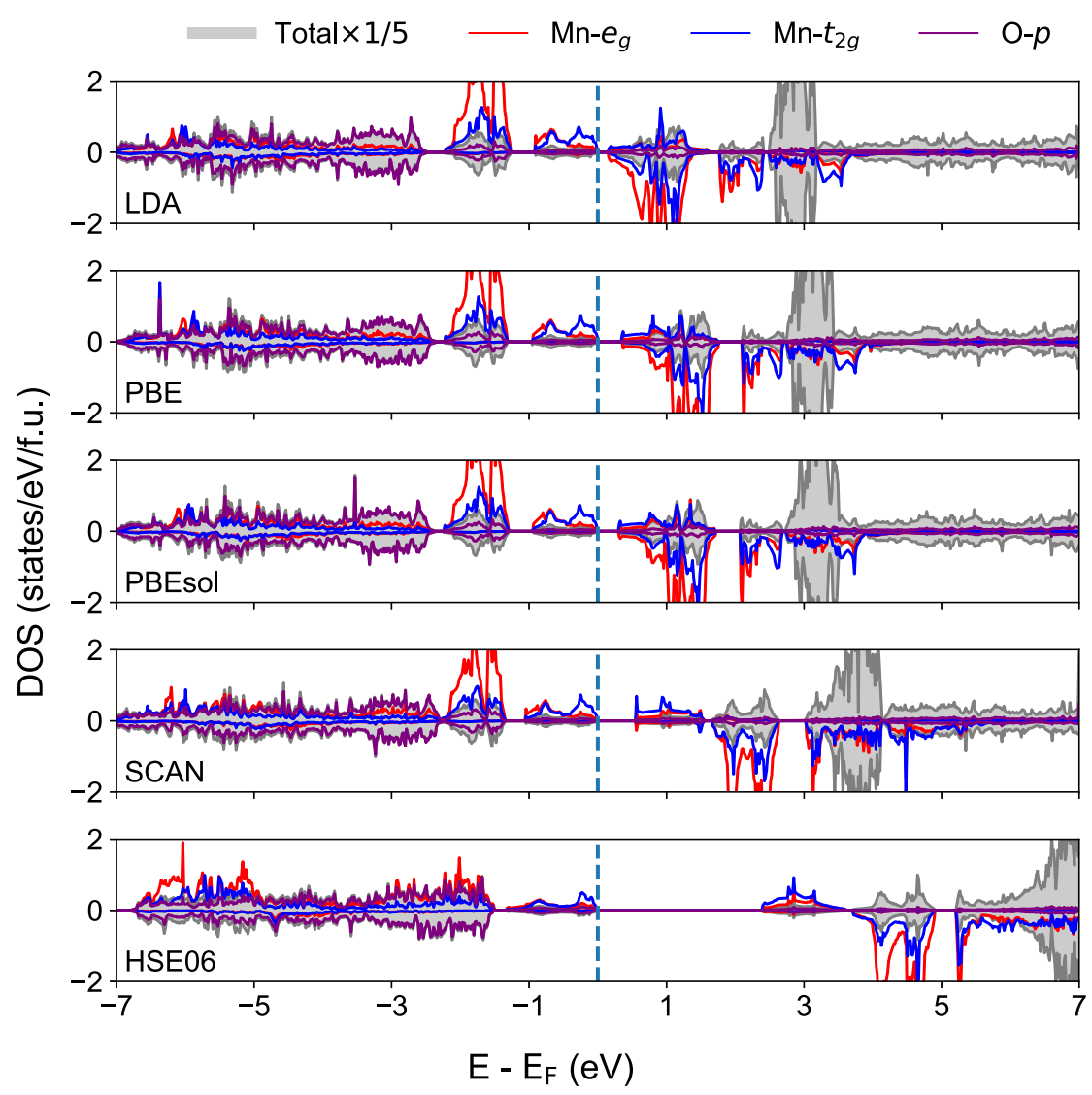

**Fig. 8.** Projected density of states calculated using the LDA, PBE, PBEsol, SCAN, and HSE06 functionals. Red and blue lines represent the Mn-$e_g$ and $t_{2g}$ orbitals, respectively, while the O-$p$ orbital is shown in purple. The total density of states is indicated by gray shading and scaled by a factor of 1/5 for clarity.

Hybrid functionals are alternative approaches beyond standard DFT, mixing exact nonlocal exchange with local DFT functionals. Such methods have been extensively applied to transition metal oxides [33]. For LMO, previous studies using hybrid functionals report partial success in reproducing both the magnetic structure and the band gap [33,34]. In particular, the HSE06 functional, one of the representative hybrid functionals, corrects DFT delocalization errors by properly mixing exact Hartree–Fock exchange, defined by the mixing parameter ($\alpha$). The typical value of $\alpha$ is 0.25, and with this mixing, we observe a very strong spin splitting, as shown in Fig. 8, which corresponds to a $U$ value of 7.0 eV or more within the CDFT + $U$ scheme. Previous systematic studies have shown that the optimal $\alpha$ value is highly system dependent, even among different manganite oxides [35]. For LMO, the band gap increases linearly with $\alpha$, as shown in Fig. 9. The optimal range for $\alpha$ is found to be 0.10–0.15, corresponding to $U$ values of 4–5 eV. Compared to DFT + $U$ approaches, the HSE functional significantly modifies the unoccupied bands.

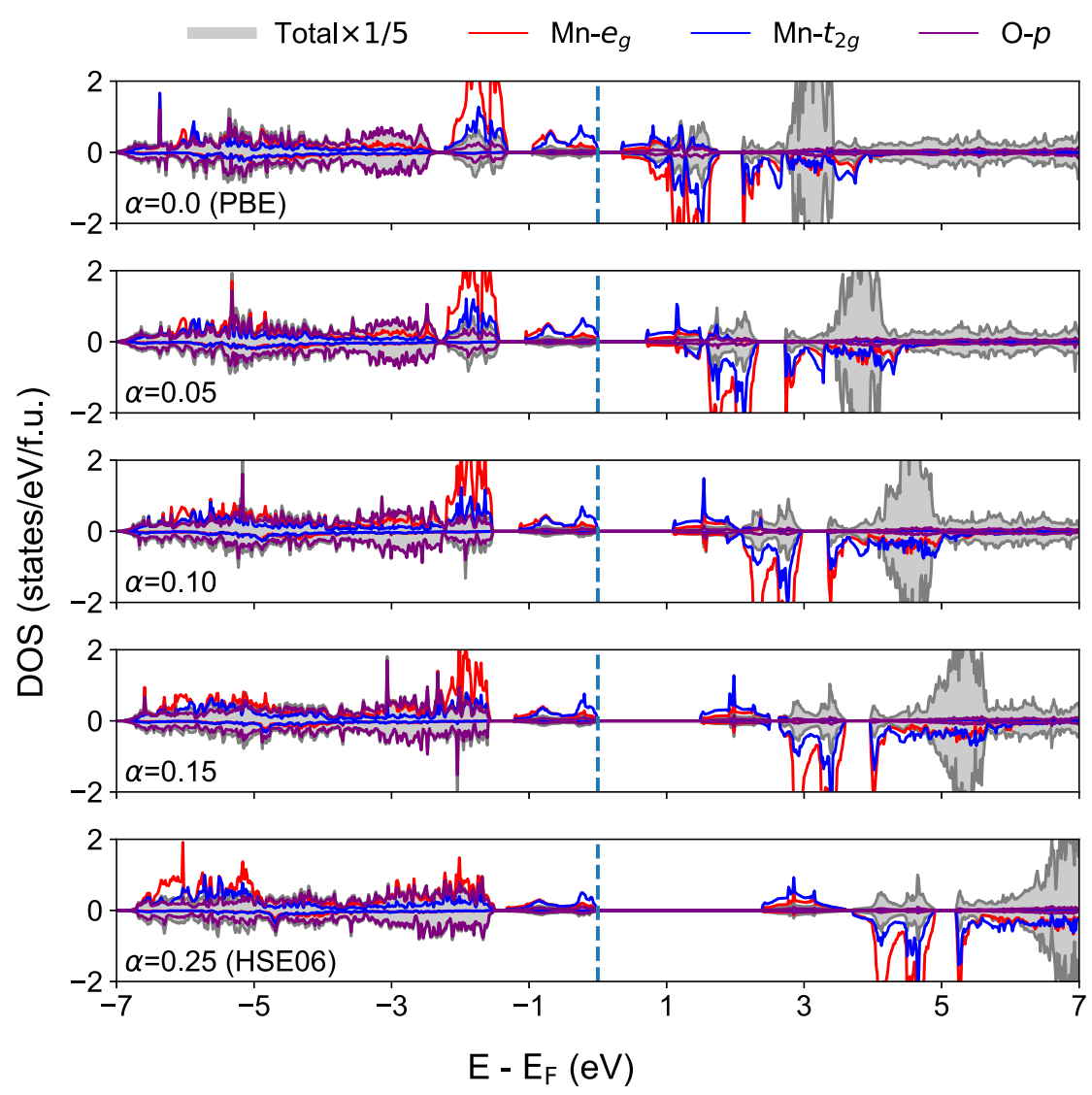

**Fig. 9.** Projected density of states calculated using the HSE functional with varying mixing parameter $\alpha$, which denotes the ratio of exact short-range exchange energy. The color coding is the same as in Fig. 8.

## 4. Conclusions

In this study, we investigated the electronic and magnetic properties of LMO using various implementations of the DFT + $U$ framework. Specifically, we compared the SDFT + $U$ and CDFT + $U$ schemes in analyzing the electronic structure and magnetic phases. Our results reveal fundamental differences between the two approaches, particularly in the treatment of Hund's $J_H$ and its explicit influence on the magnetic phase stability. The role of $J_H$ is further highlighted by mapping the magnetic exchange onto a Heisenberg-type Hamiltonian, where the CDFT + $U$ method more accurately captures the realistic parameter space of the system. We also assessed the performance of the SCAN and HSE functionals as alternatives to DFT + $U$. The limitations of the SCAN functional are evident, especially in the description of the magnetic properties.

Overall, our study highlights the challenges in accurately describing strongly correlated systems such as LMO within DFT-based frameworks. The findings emphasize the nontrivial nature of these systems and provide valuable insight for future computational studies of related materials.

Acknowledgments

This work was supported by Kyungpook National University Research Fund, 2023.